\documentclass[preprint,12pt,authoryear]{elsarticle}

\usepackage{amssymb, amsfonts, amsmath, bm}
\usepackage{booktabs} 
\usepackage{tabularx}
\usepackage{caption}
\usepackage{subcaption}
\usepackage{multirow, multicol}
\usepackage{soul}
\usepackage{array}

\usepackage{xcolor}
\usepackage{float}
\usepackage{multicol,multirow}

\journal{Computers \& Chemical Engineering}

\begin{document}

\begin{frontmatter}



\title{Augmenting optimization-based molecular design with graph neural networks}


\author[inst1]{Shiqiang Zhang}
\author[inst1]{Juan S. Campos}
\author[inst2]{Christian Feldmann}
\author[inst2]{Frederik Sandfort}
\author[inst2]{Miriam Mathea}
\author[inst1]{Ruth Misener}

\affiliation[inst1]{organization={Imperial College London},
            city={London},
           % postcode={SW7 2AZ}, 
            country={UK}}

\affiliation[inst2]{organization={BASF SE},
            city={Ludwigshafen}, 
            country={Germany}}

\begin{abstract}
Computer-aided molecular design (CAMD) studies quantitative structure-property relationships and discovers desired molecules using optimization algorithms. With the emergence of machine learning models, CAMD score functions may be replaced by various surrogates to automatically learn the structure-property relationships. Due to their outstanding performance on graph domains, graph neural networks (GNNs) have recently appeared frequently in CAMD. But using GNNs introduces new optimization challenges. This paper formulates GNNs using mixed-integer programming and then integrates this GNN formulation into the optimization and machine learning toolkit OMLT. To characterize and formulate molecules, we inherit the well-established mixed-integer optimization formulation for CAMD and propose symmetry-breaking constraints to remove symmetric solutions caused by graph isomorphism. In two case studies, we investigate fragment-based odorant molecular design with more practical requirements to test the compatibility and performance of our approaches.

\end{abstract}

\begin{keyword}
Optimization formulations \sep Molecular design \sep Graph neural networks \sep Inverse problem \sep Software tools
% %% MSC codes here, in the form: \MSC code \sep code
% %% or \MSC[2008] code \sep code (2000 is the default)
\end{keyword}

\end{frontmatter}


\section{Introduction}\label{sec:intro}
Computer-aided molecular design (CAMD)  uses modeling and optimization algorithms to discover and develop new molecules with desired properties \citep{Gani2004,Ng2014,Austin2016,Chong2022,Gani2022,Mann2023}. Recent CAMD advances use the rapid development of machine learning to create surrogate models which learn structure-property relationships and can then score and/or generate molecules \citep{Elton2019,Alshehri2020,Alshehri2021,Faez2021,Gao2022,Hatamleh2022,Tiew2023}. CAMD also heavily relies on numerical optimization algorithms to find reasonable and promising candidates in a specific design space: the resulting optimization problems are typically nonconvex and nonlinear \citep{Camarda1999,Ng2014,Zhang2015}.

Due to the natural graph representation of molecules, graph neural networks (GNNs) are attractive in CAMD \citep{Xia2019,Yang2019,Xiong2021,Rittig2022}. GNNs can predict molecular properties given the graph structure of molecules \citep{Gilmer2017,Xu2017,Shindo2019,Wang2019,Schweidtmann2020,Withnall2020}. Molecular design in these GNN-based approaches requires optimizing over graph domains. 
Most works use optimization algorithms such as genetic algorithms and Bayesian optimization that \emph{evaluate} GNNs rather than directly handling the inverse problems defined on GNNs \citep{Jin2018,Jin2020,Rittig2022}. In these works, GNNs are typically used as forward functions during the optimization process. One exception is the work of \citet{McDonald2023}, who first encoded both the graph structures of inputs and the inner structures of GNNs into a specific CAMD problem using mixed-integer programming.

Mixed-integer programming (MIP) can directly formulate molecules with optimization constraints. The basic idea is representing a molecule as a graph, creating variables for each atom (or group of atoms), and using constraints to preserve graph structure and satisfy chemical requirements. Such representations are well-established in the CAMD literature \citep{Odele1993,Churi1996,Camarda1999,Sinha1999,Sahinidis2003,Zhang2015,Liu2019,Cheun2023}. A score function with closed form is usually given as the optimization target. In addition to building the score function with knowledge from experts or statistics, we advocate using machine learning models, including GNNs, as score functions. But involving machine learning models brings new challenges to optimization: our work seeks to make it tractable to optimize over these models.

Mathematical optimization over trained machine learning models is an area of increasing interest. In this paradigm, a trained machine learning model is translated into an optimization formulation, thereby enabling decision-making problems over model predictions. For continuous, differentiable models, these problems can be solved via gradient-based methods \citep{Szegedy2014,Bunel2020a,Wu2020b,Horvath2021}. MIP has also been proposed, mainly to support non-smooth models such as ReLU neural networks \citep{Fischetti2018,Anderson2020,Tsay2021}, their ensembles \citep{Wang2023}, and tree ensembles \citep{Misic2020,Mistry2021,Thebelt2021}. 
Optimization over trained ReLU neural networks has been especially prominent \citep{Huchette2023}, finding applications such as verification \citep{Bunel2018,Tjeng2019,Botoeva2020,Bunel2020b}, reinforcement learning \citep{Say2017,Delarue2020,Ryu2020}, compression \citep{Serra2021}, and black-box optimization \citep{Papalexopoulos2022}.

Motivated by the intersection of MIP, GNNs, and CAMD, our conference paper \citep{Zhang2023} optimizes over trained GNNs with a molecular design application and combines mixed-integer formulations of GNNs and CAMD. By breaking symmetry caused by graph isomorphism, we reduced the redundancy in the search space and sped up the solving process. This paper extends this conference paper. Section \ref{sec:optimization_GNN} and Section \ref{sec:MIP_CAMD} primarily derive from \citet{Zhang2023}.
The new contributions of this paper include:
\begin{itemize}
    \item Improving the mixed-integer formulation of GNNs with tighter constraints in Section \ref{subsec:MIP_GNN}.
    \item Implementing the GNN encoding into the optimization and machine learning toolkit OMLT \citep{Ceccon2022}.
    \item Extending the mixed-integer formulation of CAMD from atom-based to fragment-based design in Section \ref{sec:case_studies}, this change admits larger molecules with aromacity.
    \item Investigating two case studies in Section \ref{subsec:banana_odor} and Section \ref{subsec:garlic_odor}. We incorporate several practical requirements into our formulations to design more realistic molecules. 
\end{itemize}

\textbf{Paper structure} Section \ref{sec:optimization_GNN} introduces optimization over trained GNNs. Section \ref{sec:GNN_OMLT} discusses the implementation of GNNs in OMLT. Section \ref{sec:MIP_CAMD} presents the mixed-integer formulation for CAMD and symmetry-breaking constraints. Section \ref{sec:case_studies} shows two case studies with specific requirements. Section \ref{sec:numerical_results} provides numerical results. Section \ref{sec:conclusion} concludes and discusses future work.

\section{Optimization over trained graph neural networks}\label{sec:optimization_GNN}
\subsection{Definition of graph neural networks}\label{subsec:GNN_definition}
This work considers a GNN with $L$ layers:
    \begin{align*}
        GNN:\underbrace{\mathbb R^{d_0}\otimes\cdots\otimes\mathbb R^{d_0}}_{|V|\ \rm{times}}\to\underbrace{\mathbb R^{d_L}\otimes\cdots\otimes\mathbb R^{d_L}}_{|V|\ \rm{times}},
    \end{align*}
where $V$ is the set of nodes of the input graph. Let ${\bm x}_v^{(0)} \in \mathbb{R}^{d_0}$ be the input features for node $v$. Then, the $l$-th layer ($l=1,2,\dots,L$) is defined by:
    \begin{align}\tag{$l^{\text{th}}$ layer}\label{def of lth layer}
        {\bm x}_v^{(l)}=\sigma\left(\sum\limits_{u\in\mathcal N(v)\cup\{v\}}{\bm w}_{u\to v}^{(l)}{\bm x}_u^{(l-1)}+{\bm b}_{v}^{(l)}\right),~\forall v\in V,
    \end{align}
where $\mathcal N(v)$ is the set of all neighbors of $v$ and $\sigma$ is any activation function.

With linear aggregate functions such as sum and mean, many classic GNN architectures can be rewritten in form \eqref{def of lth layer}, e.g., Spectral Network \citep{Bruna2014}, Neural FPs \citep{Duvenaud2015}, DCNN \citep{Atwood2016}, ChebNet \citep{Defferrard2016}, PATCHY-SAN \citep{Niepert2016}, MPNN \citep{Gilmer2017}, GraphSAGE \citep{Hamilton2017}, and GCN \citep{Kipf2017}. 

\subsection{Problem definition}\label{subsec:probelm_definition}
Given a trained GNN in form \eqref{def of lth layer}, we consider optimization problems defined as:
    \begin{align*} 
        \min\limits_{({\bm x}_v^{(0)},\dots,{\bm x}_v^{(L)},A)}~&obj({\bm x}_v^{(0)},\dots,{\bm x}_v^{(L)},A) \label{OPT}\tag{OPT}\\
        s.t.~&f_i({\bm x}_v^{(0)},\dots,{\bm x}_v^{(L)},A)\le 0,i\in \mathcal I,\\
        ~&g_j({\bm x}_v^{(0)},\dots,{\bm x}_v^{(L)},A)=0,j\in\mathcal J,\\
        ~&\eqref{def of lth layer}, 1\le l\le L,
    \end{align*}
where $A$ is the adjacency matrix of input graph, $f_i,g_j$ are problem-specific constraints and $\mathcal I,\mathcal J$ are index sets.    

\subsection{Mixed-integer optimization formulations for graph neural networks}\label{subsec:MIP_GNN}
If the graph structure for inputs is given and fixed, then a GNN layer in form \eqref{def of lth layer} is equivalent to a dense, i.e., fully connected, layer, whose MIP formulations are established \citep{Anderson2020,Tsay2021}. But if the graph structure is non-fixed, the elements in adjacency matrix $A$ are also decision variables. 

Assuming that the weights and biases are constant, \citet{Zhang2023} developed both a bilinear and a big-M formulation. The bilinear formulation results in a mixed-integer quadratically constrained optimization problem, which can be handled by state-of-the-art solvers. The \citet{Zhang2023} big-M formulation generalizes \citet{McDonald2023} from GraphSAGE \citep{Hamilton2017} to all GNN architectures satisfying \eqref{def of lth layer}. %The \citet{Zhang2023} numerical results show that the big-M formulation outperforms the bilinear one in both solving time and the earliest time finding optimum, so this paper only introduces the big-M formulation.
The \citet{Zhang2023} numerical results show that the big-M formulation outperforms the bilinear one in both (i) the time when the optimal solution was found and (ii) the time spend proving that this solution is optimal, so this paper only introduces the big-M formulation.

The big-M formulation adds auxiliary variables ${\bm z}_{u\to v}^{(l-1)}$ to represent the contribution from node $u$ to node $v$ in the $l$-th layer:
    \begin{align}\tag{big-M}\label{big-M formulation}
        {\bm x}_v^{(l)}=\sigma\left(\sum\limits_{u\in V}{\bm w}_{u\to v}^{(l)}{\bm z}_{u\to v}^{(l-1)}+{\bm b}_{v}^{(l)}\right),~\forall v\in V,
    \end{align}
where ${\bm z}_{u\to v}^{(l-1)}=A_{u,v}{\bm x}_u^{(l-1)}$ is constrained using big-M:
    \begin{align*}
        {\bm x}_{u}^{(l-1)}-{\bm U}_{u}^{(l-1)}(1-A_{u,v})\le &{\bm z}_{u\to v}^{(l-1)}\le {\bm x}_{u}^{(l-1)}-{\bm L}_{u}^{(l-1)}(1-A_{u,v}),\\
        {\bm L}_{u}^{(l-1)}A_{u,v}\le &{\bm z}_{u\to v}^{(l-1)}\le {\bm U}_u^{(l-1)}A_{u,v},
    \end{align*}
where ${\bm L}_u^{(l-1)}\le {\bm x}_u^{(l-1)}\le {\bm U}_u^{(l-1)}, A_{u,v}\in\{0,1\}$. Given the bounds of input features $\{{\bm x}_v^{(0)}\}_{v\in V}$, all bounds of $\{{\bm x}_v^{(l)}\}_{v\in V,1\le l\le L}$ can be derived using interval arithmetic. Note that these big-M constraints provide tighter bounds for ${\bm z}_{u\to v}^{(l-1)}$ than the original \citet{Zhang2023} formulation.

\section{Encoding graph neural networks into OMLT}\label{sec:GNN_OMLT}
OMLT \citep{Ceccon2022} is an open-source software package that encodes trained machine learning models into the algebraic modeling language Pyomo \citep{Bynum2021}. Taking a trained dense neural network as an example, by defining variables for all neurons and adding constraints to represent links between layers, OMLT automatically transforms the neural network into a Pyomo block suitable for optimization. Currently, OMLT supports dense neural networks \citep{Anderson2020, Tsay2021}, convolutional neural networks \citep{Albawi2017},  gradient-boosted trees \citep{Thebelt2021}, and linear model decision trees \citep{Ammari2023}.

\begin{figure}[t]
        \centering
        \includegraphics[width=0.9\textwidth]{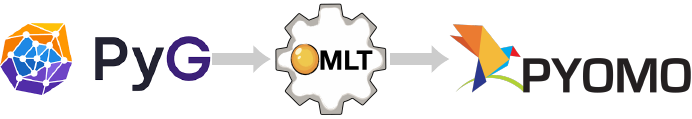}
        \caption{With the contributions in this paper, OMLT now transforms a PyG GNN sequential model into a Pyomo block.}
        \label{fig:PyG_OMLT_Pyomo}
    \end{figure}

To enable GNNs in OMLT, we create a \textbf{GNNLayer} class defined by \eqref{def of lth layer} and encode this class using the Section \ref{subsec:MIP_GNN} big-M formulation. As shown in Figure \ref{fig:PyG_OMLT_Pyomo}, our extension to OMLT now supports GNNs through a Pytorch Geometric PyG \citep{Fey2019} interface. Note that OMLT requires a sequential GNN model in PyG as a standard format.

In this section, we first describe the \textbf{GNNLayer} class and provide some examples to show how to transform various GNN-relevant operations to OMLT in Section \ref{subsec:GNNLayer_class}. Section \ref{subsec:GNN_formulation} shows details about encoding \textbf{GNNLayer}. Section \ref{subsec:PyG_OMLT} lists all currently supported operations in our extension to OMLT.

\subsection{GNNLayer class in OMLT}\label{subsec:GNNLayer_class}
The OMLT \textbf{GNNLayer} class requires the following information:
    \begin{itemize}
        \item $N$: number of nodes.
        \item Input/Output size: number of input/output features for all nodes. Note that the input/output size should be equal to $N$ multiplied by the number of input/output features for each node.
        \item Weight matrix $\bm W$: consists of all weights ${\bm w}_{u\to v},~\forall u,v\in [N]$.
        \item Biases $\bm B$: consists of all biases ${\bm b}_{v},~\forall v\in [N]$.
    \end{itemize}
    where $[N]:=\{0,1,\dots,N-1\}$.

As long as a GNN layer satisfies definition \eqref{def of lth layer}, it could be defined in the \textbf{GNNLayer} class and encoded using \eqref{big-M formulation}. The extra effort is that users need to properly transform their GNN layers to OMLT. We provide several examples to illustrate how to define a GNN layer inside OMLT. 

\textbf{Remark: }This section only considers a single layer and drops the indexes of layers. To differentiate the input and output features, we use ${\bm y}_{v}$ as the output features of node $v$.

\noindent
\textbf{Example 1 [Linear]} Consider a linear layer:
    \begin{align*}
        {\bm y}_v = {\bm w}^T{\bm x}_v+{\bm b}, ~\forall v\in [N].
    \end{align*}
This is different from a common dense layer since it works on features of each node separately. 
%This layer could be transformed into a dense layer in OMLT with parameters:
Our new extension to OMLT transforms this layer into a dense layer with parameters:
    \begin{align*}
        {\bm W}=
        \begin{pmatrix}
            \bm w &  &  &  \\
             & \bm w & & \\
            & & \ddots & \\
            & & & \bm w
        \end{pmatrix},
        ~{\bm B}=
        \begin{pmatrix}
            {\bm b}\\
            {\bm b}\\
            \vdots\\
            {\bm b}
        \end{pmatrix}.
    \end{align*}

\noindent
\textbf{Example 2 [GCNConv with fixed graph]} Given a GCNConv layer:
    \begin{align*}
        {\bm y}_v={\bm w}^T\sum\limits_{u\in\mathcal N(v)\cup\{v\}}\frac{A_{u,v}}{\sqrt{\hat d_{u}\hat d_{v}}}{\bm x}_u+{\bm b},~\forall v\in [N],
    \end{align*}
where $\hat d_v=1+\sum\limits_{u\in\mathcal N(v)}A_{u,v}$. Our new extension to OMLT defines a GNN layer in OMLT with parameters:
    \begin{align*}
        {\bm W}=
        \begin{pmatrix}
            \frac{A_{0,0}}{\sqrt{\hat d_{0}\hat d_{0}}}{\bm w} & \frac{A_{0,1}}{\sqrt{\hat d_{0}\hat d_{1}}}{\bm w} & \cdots & \frac{A_{0,N-1}}{\sqrt{\hat d_{0}\hat d_{N-1}}}{\bm w} \\
            \frac{A_{1,0}}{\sqrt{\hat d_{1}\hat d_{0}}}{\bm w} & \frac{A_{1,1}}{\sqrt{\hat d_{1}\hat d_{1}}}{\bm w} & \cdots & \frac{A_{1,N-1}}{\sqrt{\hat d_{1}\hat d_{N-1}}}{\bm w} \\
            \vdots & \vdots & \ddots & \vdots \\
            \frac{A_{N-1,0}}{\sqrt{\hat d_{N-1}\hat d_{0}}}{\bm w} & \frac{A_{N-1,1}}{\sqrt{\hat d_{N-1}\hat d_{1}}}{\bm w} & \cdots & \frac{A_{N-1,N-1}}{\sqrt{\hat d_{N-1}\hat d_{N-1}}}{\bm w}
        \end{pmatrix},
        ~{\bm B}=
        \begin{pmatrix}
            {\bm b}\\
            {\bm b}\\
            \vdots\\
            {\bm b}
        \end{pmatrix}.
    \end{align*}

Since $A$ is given and fixed, all coefficients $\frac{A_{u,v}}{\sqrt{\hat d_{u}\hat d_{v}}}$ are constant. $\bm w,\bm b$ are also constant for a trained GNN. Therefore, $\bm W, \bm B$ are constant. However, when $A$ is not fixed, a GCNConv layer results in the nonlinear formulation:
    \begin{align*}
        {\bm y}_v=\sum\limits_{u\in [N]}\frac{A_{u,v}}{\sqrt{(1+\sum\limits_{w\in\mathcal N(u)}A_{w,u})(1+\sum\limits_{w\in\mathcal N(v)}A_{w,v})}}{\bm w}^T{\bm x}_u+{\bm b},~\forall v\in [N].
    \end{align*}

\noindent
\textbf{Example 3 [SAGEConv with non-fixed graph]} Given a SAGEConv layer with sum aggregation defined by:
\begin{align*}
    {\bm y}_v={\bm w}^T_1{\bm x}_v+{\bm w}_2^T\sum\limits_{u\in\mathcal N(v)}{\bm x}_u+{\bm b},~\forall v\in [N].
\end{align*}
Since the graph structure is not fixed, we need to provide all possibly needed weights and biases, i.e., assume a complete graph, and then define binary variables $A_{u,v}$ required in \eqref{big-M formulation}. The corresponding GNN layer in our new extension to OMLT is defined with parameters:
\begin{align*}
    {\bm W}=
    \begin{pmatrix}
        {\bm w}_1 & {\bm w}_2 & \cdots & {\bm w}_2 \\
        {\bm w}_2 & {\bm w}_1 & \cdots & {\bm w}_2 \\
        \vdots & \vdots & \ddots & \vdots \\
        {\bm w}_2 & {\bm w}_2 & \cdots & {\bm w}_1
    \end{pmatrix},
    ~{\bm B}=
    \begin{pmatrix}
        {\bm b}\\
        {\bm b}\\
        \vdots\\
        {\bm b}
    \end{pmatrix}.
\end{align*}

\noindent
\textbf{Example 4 [global\_mean\_pooling]} A pooling layer gathers features of all nodes to form a graph representation, for instance, a mean pooling layer is defined by:
\begin{align*}
    {\bm y}=\frac{1}{N}\sum\limits_{v\in [N]}{\bm x}_v,
\end{align*}
which is transformed into a dense layer in our new OMLT extension with parameters:
\begin{align*}
    {\bm W}=
    \begin{pmatrix}
        \frac{1}{N} & \cdots & \frac{1}{N} & & & & & & &\\
        & & & \frac{1}{N} & \cdots & \frac{1}{N} & & & &\\
        & & & & & & \ddots & & & \\ 
        & & & & & & & \frac{1}{N} & \cdots & \frac{1}{N} 
    \end{pmatrix}, ~{\bm B}={\bm 0}.
\end{align*}
The size of ${\bm W}$ is $F\times NF$, where $F$ is the length of ${\bm x}_v$. 

\subsection{Formulating GNNLayer in OMLT}\label{subsec:GNN_formulation}
OMLT formulates neural networks layer by layer. Each layer class is associated with a specific formulation that defines variables for neurons and links these variables using constraints. By gathering all variables and constraints into a Pyomo block, OMLT encodes a neural network into an optimization problem. All problem-specific variables and constraints can also be included into the block to represent practical requirements.

With the new \textbf{GNNLayer} class in OMLT, the last step is encoding the \eqref{big-M formulation} formulation in Section \ref{subsec:MIP_GNN} and associating it with \textbf{GNNLayer} by:
\begin{enumerate}
    \item Defining binary variables $A_{u,v},u,v\in[N]$ for all elements in the adjacency matrix $A$. Note: these variables are shared by all GNN layers.
    \item defining variables for all features ${\bm x}_v^{(l)},v\in [N],0\le l\le L$.
    \item Defining auxiliary variables ${\bm z}_{u\to v}^{(l-1)},u,v\in[N],1\le l\le L$.
    \item Defining bounds $\left[{\bm L}_{u\to v}^{(l-1)},{\bm U}_{u\to v}^{(l-1)}\right]$ for ${\bm z}_{u\to v}^{(l-1)},u,v\in[N],1\le l\le L$ as:
        \begin{align*}
            \begin{cases}
                \left[{\bm 0},{\bm 0}\right], & \text{$A_{u,v}$ is fixed to $0$}\\
                \left[{\bm L}_{u}^{(l-1)},{\bm U}_{u}^{(l-1)}\right], & \text{$A_{u,v}$ is fixed to $1$}\\
                \left[\min\left({\bm 0},{\bm L}_{u}^{(l-1)}\right),\max\left({\bm 0},{\bm U}_{u}^{(l-1)}\right)\right], & \text{$A_{u,v}$ is not fixed}\\
            \end{cases}
        \end{align*}
    where $\left[{\bm L}_{u}^{(l-1)},{\bm U}_{u}^{(l-1)}\right]$ are bounds of ${\bm x}_u^{(l-1)}$.
    \item Adding constraints to represent \eqref{big-M formulation}.
\end{enumerate}

After formulating the \textbf{GNNLayer} class, we finish the implementation of GNNs in OMLT. For simplicity of notations, we only consider a GNN defined as \eqref{def of lth layer}. Benefiting from the layer-based architecture of OMLT, all layers that OMLT currently supports are compatible with the GNN encoding, as long as the outputs of the previous layer match the inputs of the next layer. For instance, OMLT can encode a GNN consisting of several GNN layers, a pooling layer, and several dense layers.

\subsection{Supported PyG operations in OMLT}\label{subsec:PyG_OMLT}
This section discusses our implementation of PyG operations in OMLT to facilitate GNN encoding. Table \ref{table:PyG_operations} summarizes all implemented operations with fixed or non-fixed graph structure. By assembling a GNN sequential model consisting of these operations, users may encode GNNs into OMLT without bothering with the complex transformations from Section \ref{subsec:GNNLayer_class}. 

\begin{table}[h]
    \centering
    \caption{Supported PyG operations in OMLT.}
    \label{table:PyG_operations}
    \begin{tabular}{llcc}
    \toprule
    \multirow{2}{*}{Type} & \multirow{2}{*}{Operations} & \multicolumn{2}{c}{Graph structure}\\
    & & fixed & non-fixed \\
    \midrule
    \multirow{3}{*}{Convolution} & \textbf{Linear} & $\surd$ & $\surd$ \\
    & \textbf{GCNConv} & $\surd$ & $\times$ \\
    & \textbf{SAGEConv} & $\surd$ & $\surd$ \\
    \midrule
    \multirow{2}{*}{Aggregation} & \textbf{mean} & $\surd$ & $\times$ \\
    & \textbf{sum} & $\surd$ & $\surd$ \\
    \midrule
    \multirow{2}{*}{Pooling} & \textbf{global\_mean\_pooling} & $\surd$ & $\surd$ \\
    & \textbf{global\_add\_pooling} & $\surd$ & $\surd$ \\
    \bottomrule
    \end{tabular}
\end{table}

As discussed in Section \ref{subsec:MIP_GNN}, when the input graph structure is not fixed, our extension of OMLT assumes constant weights and biases to formulate a GNN layer. The cost for this assumption is that \textbf{GCNConv} and \textbf{mean} aggregation are not supported in this case since both result in difficult mixed-integer nonlinear optimization problems, e.g., as shown in Example 2. 

In OMLT, activation functions are detached from layers. Therefore, all activation functions in OMLT are compatible with GNNs, including \textbf{ReLU}, \textbf{Sigmoid}, \textbf{LogSoftmax}, \textbf{Softplus}, and \textbf{Tanh}. Among these activations, ReLU can be represented by linear constraints using a big-M formulation \citep{Anderson2020}. Because of the binary variables $A_{u,v}$ representing the existence or not of an edge (see Section \ref{subsec:MIP_GNN}), any 
activation function formulated with nonlinear constraints will result in a mixed-integer non-linear optimization problem. Thus, although OMLT will allow nonlinear activation functions for GNNs, practical algorithmic limitations may make nonlinear activations functions inadvisable in OMLT.

It is noteworthy that importing a GNN model from PyG interface is not the only way to encode GNNs in OMLT. One can transform their own GNNs into OMLT in a similar way to those examples in Section \ref{subsec:GNNLayer_class}. As long as one can properly define their GNN operations into \textbf{GNNLayer} class, OMLT will formulate these layers using \ref{big-M formulation} automatically.

\section{Mixed-integer optimization formulation for molecular design}\label{sec:MIP_CAMD}
MIP formulations are well-established in the CAMD literature. Our particular MIP formulation for CAMD starts with that of \cite{McDonald2023}. Because our formulation also incorporates Sections \ref{subsec:proper_bounds} and \ref{subsec:symmetry_breaking_constraints}, we are able to develop optimization approaches that find larger molecules (with $12$ heavy atoms or fragments rather than $4$).

\subsection{Atom features}\label{subsec:atom_features}
To design a molecule with at most $N$ atoms, we define $N\times F$ binary variables $X_{v,f},v\in [N],f\in [F]$ to represent $F$ features for $N$ atoms. Hydrogen atoms are not counted in since they can implicitly be considered as node features. Atom features include types of atom, number of neighbors, number of hydrogen atoms associated, and types of adjacent bonds. Table \ref{table:summary_of_notations} summarizes relevant notations and provides their values for an example with four types of atoms $\{C,N,O,S\}$. Table \ref{table:atom_features} describes atom features explicitly.

    \begin{table}[bp] 
        \caption{Summary of notations used in MIP formulation for CAMD. Their values correspond to the example with four types of atom $\{C,N,O,S\}$.}
        \label{table:summary_of_notations}
        \centering
        \vspace{1mm}
        \begin{tabular}{ccc}
            \toprule   
            Symbol & Description & Value \\
            \midrule
            $N$ & number of nodes &  *\\
            $F$ & number of features & $15$ \\
            $N^t$ & number of atom types & $4$ \\
            $N^n$ & number of neighbors & $4$ \\
            $N^h$ & number of hydrogen & $5$ \\
            \midrule
            $I^t$ & indexes for $N^t$ & $\{0,1,2,3\}$ \\
            $I^n$ & indexes for $N^n$ & $\{4,5,6,7\}$ \\
            $I^h$ & indexes for $N^h$ & $\{8,9,10,11,12\}$ \\
            $I^{db}$ & index for double bond & $13$ \\
            $I^{tb}$ & index for triple bond & $14$ \\
            \midrule
            $Atom$ & atom types &  $\{C,N,O,S\}$ \\ 
            $Cov$ & covalences of atom & $\{4,3,2,2\}$ \\
            \bottomrule  
        \end{tabular}
    \end{table}

    \begin{table}[bp] 
        \caption{List of atom features}
        \label{table:atom_features}
        \centering
        \vspace{1mm}
        \begin{tabular}{cc}
            \toprule
            $X_{v,f}$ & Description \\
            \midrule
            $f\in I^t$ & which atom type in $Atom$ \\
            $f\in I^n$ & number of neighbors, $1\sim \max\{Cov\}$ \\
            $f\in I^h$ & number of hydrogen, $0\sim \max\{Cov\}$ \\
            $f=I^{db}$ & if $v$ is included with double bond(s) \\
            $f=I^{tb}$ & if $v$ is included with triple bond \\
            \bottomrule
        \end{tabular}
        
    \end{table}

\subsection{Bond features}\label{subsec:bond_features}
For bond features, we add three sets of binary variables $A_{u,v},DB_{u,v},TB_{u,v}$ to denote any bond, double bond, and triple bond between atom $u$ and atom $v$, respectively:
    \begin{itemize}
        \item $A_{u,v},u\neq v\in [N]$: if there is a bond between atom $u$ and $v$.
        \item $A_{v,v},v\in [N]$: if node $v$ exists.
        \item $DB_{u,v},u\neq v\in [N]$: if there is a double bond between atom $u$ and $v$.
        \item $TB_{u,v},u\neq v\in [N]$: if there is a triple bond between atom $u$ and $v$.
    \end{itemize}

\textbf{Remark: }$\{A_{v,v}\}_{v\in [N]}$ allow us to design molecules with at most $N$ atoms. For comparison purposes in our experiments, however, we set $A_{v,v}=1,~\forall v\in [N]$ to design molecules with exactly $N$ atoms.

\subsection{Structural constraints}\label{subsec:structural_constraints}
This section introduces constraints to handle structural feasibility, which is commonly considered in the literature \citep{Odele1993,Churi1996,Camarda1999,Sinha1999,Sahinidis2003,Zhang2015}. 

\subsubsection{Feasible adjacency matrix}
Constraints \eqref{C1} represent the minimal requirement, i.e., there are two atoms and one bond between them. \eqref{C2} force that atoms with smaller indexes exist. \eqref{C3} require symmetric $A$. \eqref{C4} only admit bonds between two existing atoms, where $N-1$ is a big-M constant. \eqref{C5} force atom $v$ to be linked with at least one atom with smaller index if atom $v$ exists.
{\allowdisplaybreaks
    \begin{align*}
        A_{0,0}=A_{1,1}=A_{0,1}&=1 &&\label{C1}\tag{C1} \\
        A_{v,v}&\ge A_{v+1,v+1},&&\forall v\in [N-1] \label{C2}\tag{C2} \\
        A_{u,v}&=A_{v,u},&&\forall u,v\in [N],u<v \label{C3}\tag{C3} \\
        (N-1)\cdot A_{v,v}&\ge \sum\limits_{u\neq v} A_{u,v},&&\forall v\in [N] \label{C4}\tag{C4} \\
        A_{v,v}&\le \sum\limits_{u<v} A_{u,v},&&\forall v\in [N]\backslash\{0\} \label{C5}\tag{C5}
    \end{align*}
    }

\subsubsection{Feasible bond features}
Constraints \eqref{C6} -- \eqref{C9} define symmetric double/triple bond matrix with zero diagonal elements. \eqref{C10} restrict bond type when this bond exists.
{\allowdisplaybreaks
    \begin{align*}
        DB_{v,v}&=0,&&\forall v\in [N] \label{C6}\tag{C6} \\
        DB_{u,v}&=DB_{v,u},&&\forall u,v\in [N],u<v \label{C7}\tag{C7} \\
        TB_{v,v}&=0,&&\forall v\in [N] \label{C8}\tag{C8} \\
        TB_{u,v}&=TB_{v,u},&&\forall u,v\in [N],u<v \label{C9}\tag{C9} \\
        DB_{u,v}+TB_{u,v}&\le A_{u,v},&&\forall u,v\in [N],u<v \label{C10}\tag{C10} 
    \end{align*}
    }

\subsubsection{Feasible atom features}
If atom $v$ exists, i.e., $A_{v,v}=1$, then constraints \eqref{C11} -- \eqref{C13} excludes more than one atom type, fixes the number of neighbors, and sets the number of associated hydrogen atoms.
{\allowdisplaybreaks
    \begin{align*}
        A_{v,v}&=\sum\limits_{f\in I^t}X_{v,f},&&\forall v\in [N] \label{C11}\tag{C11} \\
        A_{v,v}&=\sum\limits_{f\in I^n}X_{v,f},&&\forall v\in [N] \label{C12}\tag{C12} \\
        A_{v,v}&=\sum\limits_{f\in I^h}X_{v,f},&&\forall v\in [N] \label{C13}\tag{C13}
    \end{align*}
    }

\subsubsection{Compatibility between atoms and bonds}
Constraints \eqref{C14} match the number of neighbors calculated from the adjacency matrix and atom features. \eqref{C15} -- \eqref{C20} consider the compatibility between atom features and bond features. \eqref{C21} are the covalence equations.
{\allowdisplaybreaks
    \begin{align*}
        \sum\limits_{u\neq v}A_{u,v}&=\sum\limits_{i\in [N^n]}(i+1)\cdot X_{v,I^n_i},&&\forall v\in [N] \label{C14}\tag{C14} \\
        3\cdot DB_{u,v}&\le X_{u,I^{db}}+X_{v,I^{db}}+A_{u,v},&&\forall u,v\in [N],u<v \label{C15}\tag{C15} \\
        3\cdot TB_{u,v}&\le X_{u,I^{tb}}+X_{v,I^{tb}}+A_{u,v},&&\forall u,v\in [N],u<v \label{C16}\tag{C16} \\
        \sum\limits_{u\in [N]} DB_{u,v}&\le \sum\limits_{i\in[N^t]} \left\lfloor\frac{Cov_i}{2}\right\rfloor \cdot X_{v,I^t_i},&&\forall v\in [N] \label{C17}\tag{C17} \\
        \sum\limits_{u\in [N]} TB_{u,v}&\le \sum\limits_{i\in[N^t]} \left\lfloor\frac{Cov_i}{3}\right\rfloor \cdot X_{v,I^t_i},&&\forall v\in [N] \label{C18}\tag{C18} \\
        X_{v,I^{db}}&\le \sum\limits_{u\in [N]}DB_{u,v},&&\forall v\in [N] \label{C19}\tag{C19} \\
        X_{v,I^{tb}}&\le \sum\limits_{u\in [N]}TB_{u,v},&&\forall v\in [N] \label{C20}\tag{C20} \\
        \sum\limits_{i\in[N^t]} Cov_i\cdot X_{v,I^t_i}&=\sum\limits_{i\in [N^n]} (i+1)\cdot X_{v,I^n_i} \\ &+\sum\limits_{i\in [N^h]}i\cdot X_{i,I^h_i} \\
            &+\sum\limits_{u\in[N]}DB_{u,v}+\sum\limits_{u\in[N]}2\cdot TB_{u,v},&&\forall v\in [N] \label{C21}\tag{C21}
    \end{align*}
    }
    
\subsubsection{Avoiding spurious extrapolation}\label{subsec:proper_bounds}
Constraints \eqref{C22} -- \eqref{C25} bound the number of each type of atom, double/triple bonds, and rings using observations from the dataset. By setting proper bounds, we can control the composition of the molecule, and avoid extreme cases such as all atoms being set to oxygen, or a molecule with too many rings or double/triple bounds.
{\allowdisplaybreaks
    \begin{align*}
        \sum\limits_{v\in [N]}X_{v,I^t_i}&\in[LB_{Atom_i},~UB_{Atom_i}],&&\forall i\in [N^t] \label{C22}\tag{C22} \\
        \sum\limits_{v\in [N]}\sum\limits_{u<v}DB_{u,v}&\in[LB_{db},~UB_{db}] && \label{C23}\tag{C23} \\
        \sum\limits_{v\in [N]}\sum\limits_{u<v}TB_{u,v}&\in [LB_{tb},~UB_{tb}] && \label{C24}\tag{C24} \\
        \sum\limits_{v\in [N]}\sum\limits_{u<v}A_{u,v}-(N-1)&\in [LB_{ring},~UB_{ring}] && \label{C25}\tag{C25}
    \end{align*}
    }

\subsection{Symmetry-breaking constraints}\label{subsec:symmetry_breaking_constraints}
\begin{figure}[t]
        \centering
        \begin{tabular}{@{} m{0.42\textwidth} m{0.05\textwidth}<{\centering} m{0.42\textwidth}}
            \begin{subfigure}[b]{0.42\textwidth}
               \centering
               \includegraphics[width=.3\textwidth]{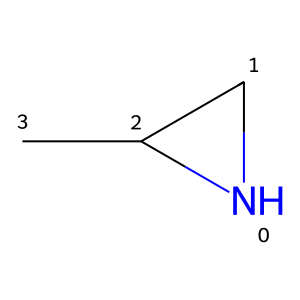}
                \includegraphics[width=.3\textwidth]{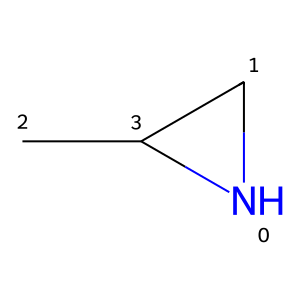}
                \includegraphics[width=.3\textwidth]{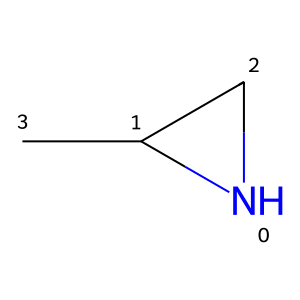}
           \end{subfigure}
           &
           $\cdots$
           &
           \begin{subfigure}[b]{0.42\textwidth}
                \centering
                \includegraphics[width=.3\textwidth]{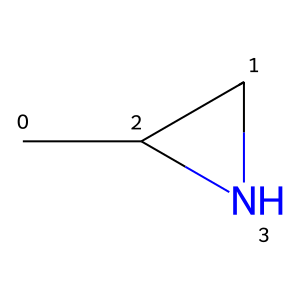}
                \includegraphics[width=.3\textwidth]{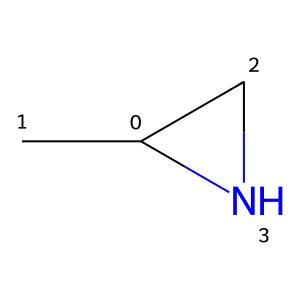}
                \includegraphics[width=.3\textwidth]{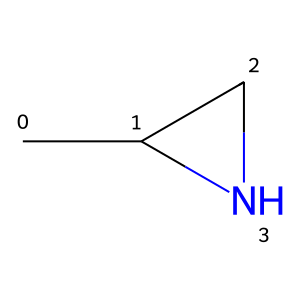}
            \end{subfigure}\\
        \end{tabular}
        \caption{A molecule (2‐methylaziridine or propylene imine) with $N=4$ heavy atoms corresponds to $N!=24$ different indexing possibilities.}
        \label{fig:symmetry_issue}
    \end{figure}

Since the properties of molecules are not influenced by the indexing of molecular graph structure, the score functions and/or models used in CAMD should be permutational invariant. Training models for CAMD benefits from this invariance since different indexing result in the same performance. However, to consider an optimization problem defined over these models, such as \eqref{OPT}, each indexing of the same molecule corresponds to a different solution. See Figure \ref{fig:symmetry_issue} as an example. These symmetric solutions significantly enlarge the search space and slow down the solving process.

\begin{figure}[t]
         \centering
         \begin{subfigure}[b]{0.32\textwidth}
             \centering
             \includegraphics[width=.4\textwidth]{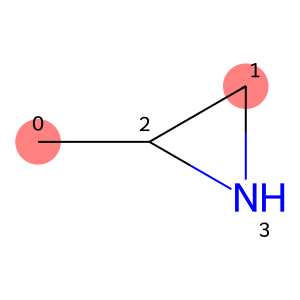}
             \caption*{\st{$(S1)$}}
         \end{subfigure}
         \hfill
         \begin{subfigure}[b]{0.32\textwidth}
             \centering
             \includegraphics[width=.4\textwidth]{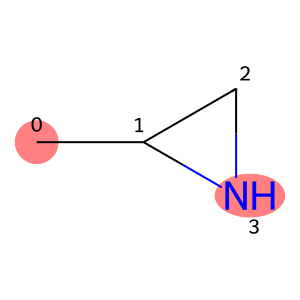}
             \caption*{\st{$(S2)$}}
         \end{subfigure}
         \hfill
         \begin{subfigure}[b]{0.32\textwidth}
             \centering
             \includegraphics[width=.4\textwidth]{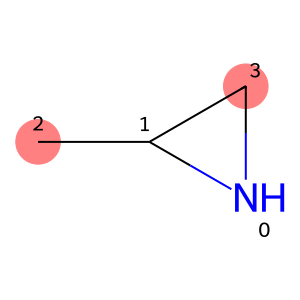}
             \caption*{\st{$(S3)$}}
         \end{subfigure}
         \caption{Applying symmetry-breaking constraints to the 2‐methylaziridine example. \st{$(S1)$}: one of the 10/24 solutions violating (S1), because node $1$ is not linked with node $0$. \st{$(S2)$}: one of the 11/13 solutions violating (S2), because the nitrogen atom is not indexed $0$. \st{$(S3)$}: one of the 2/3 solutions violating (S3), because $\mathcal N(2)=\{1\}$ but $\mathcal N(3)=\{0,1\}$.}
         \label{fig:infeasible_examples}
    \end{figure}

To handle the symmetry, \citet{Zhang2023} propose symmetry-breaking constraints. As illustrated in Figure \ref{fig:infeasible_examples}, the key ideas underlying these constraints are:

    \begin{itemize}
        \itemindent=5mm
        \item[(S1)]\label{S1} All subgraphs induced by nodes $\{0,1,\dots,v\}$ are connected.
        \item[(S2)] Node $0$ has the most special features compared to other nodes.
        \item[(S3)] Node $v$ has neighbors with smaller indexes compared to node $v+1$.
    \end{itemize}

Among these constraints, (S1) and (S3) are irrelevant to features since they focus on graph indexing. To apply (S2), an application-specific function $h$ needs to be defined to assign a hierarchy to each node based on its features. The case studies give examples of constructing $h$. Note that these constraints are not limited to CAMD: the constraints apply to any graph-based decision-making problems with symmetry issue caused by graph isomorphism.

\begin{figure}[t]
         \centering
         \begin{subfigure}[b]{0.24\textwidth}
             \centering
             \includegraphics[width=.52\textwidth]{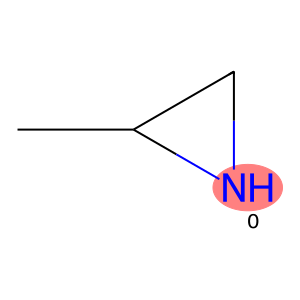}
             \caption*{Iteration 1}
         \end{subfigure}
         \begin{subfigure}[b]{0.24\textwidth}
             \centering
             \includegraphics[width=.52\textwidth]{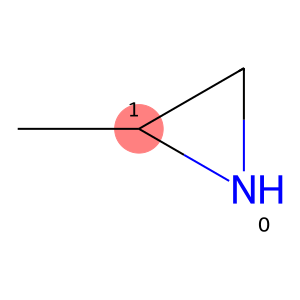}
             \caption*{Iteration 2}
         \end{subfigure}
         \begin{subfigure}[b]{0.24\textwidth}
             \centering
             \includegraphics[width=.52\textwidth]{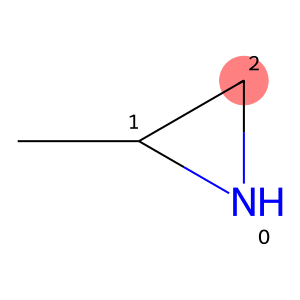}
             \caption*{Iteration 3}
         \end{subfigure}
         \begin{subfigure}[b]{0.24\textwidth}
             \centering
             \includegraphics[width=.52\textwidth]{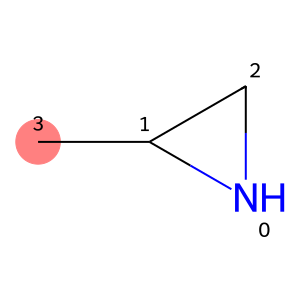}
             \caption*{Iteration 4}
         \end{subfigure}
         \caption{Using the algorithm in \citet{Zhang2023} to index 2‐methylaziridine.}
         \label{fig:indexing_example}
    \end{figure}

Among constraints \eqref{C1} - \eqref{C25}, constraints \eqref{C5} are the realization of (S1). Except for \eqref{C5}, these structural constraints are independent of the graph indexing. Therefore, we can compatibly implement constraints (S2) and (S3) to break symmetry. However, we still need to ensure that there exists at least one feasible indexing for any graph after applying these constraints. Otherwise, the diversity of the feasible set will be reduced, which is unacceptable. We proved that there exists at least one indexing satisfying both (S1) and (S3) for any graph with one node indexed $0$ \citep{Zhang2023}. If applying (S2) to choose node $0$, then we obtain the desired feasible indexing. Figure \ref{fig:indexing_example} shows how to yield a feasible indexing.

Corresponding to (S2), we add the following constraints over features:
    \begin{align}\label{C26}\tag{C26}
        \sum\limits_{f\in [F]}2^{F-f-1}\cdot X_{0,f}\le \sum\limits_{f\in [F]} 2^{F-f-1}\cdot X_{v,f}+ 2^F\cdot (1-A_{v,v}),~\forall v\in [N]\backslash\{0\},
    \end{align}
where $2^{F-f-1},f\in [F]$ are coefficients to help build a bijective $h$ between all possible features and all integers in $[0,2^F-1]$. These coefficients are also called a ``universal ordering vector" \citep{Friedman2007,Hojny2019}. Note that rearranging these coefficients will not influence the bijectivity of $h$, which will be used in our case studies. The extra term $2^F\cdot (1-A_{v,v})$ is introduced to exclude non-existant atoms.

On the graph level, constraints (S3) can equivalently be rewritten as:
    \begin{align}\label{C27}\tag{C27}
        \sum\limits_{u\neq v,v+1}2^{N-u-1}\cdot A_{u,v} \ge \sum\limits_{u\neq v,v+1}2^{N-u-1}\cdot A_{u,v+1},~\forall v\in [N-1]\backslash\{0\}.
    \end{align}
Similarly, the coefficients $2^{N-u-1},u\in [N]$ are used to build a bijective mapping between all possible sets of neighbors and all integers in $[0,2^N-1]$.

\section{Case studies}\label{sec:case_studies}
The mixed-integer optimization formulation for CAMD introduced in Section \ref{sec:MIP_CAMD} only presents the minimal requirements to molecular design. More specific constraints in real-world problems are not included. Moreover, there are two main limitations in \citet{Zhang2023}:
\begin{itemize}
    \item Problem size: we only solved the optimization problems with up to $N=8$ atoms, which are too small for most molecular design applications.
    \item Aromacity: we only considered double and triple bonds. Aromatic rings are incompatible with \eqref{C1} -- \eqref{C27}. 
\end{itemize}

To circumvent these limitations, instead of assembling atoms to form a molecule, we use the concept of molecular fragmentation \citep{Jin2018,Jin2020,Podda2020,Green2021,Powers2022}. By introducing aromatic rings as different types of fragments, we can design larger molecules with aromacity. Note that all structural constraints in Section \ref{subsec:structural_constraints} and symmetry-breaking constraints in Section \ref{subsec:symmetry_breaking_constraints} are still useful. As shown in Table \ref{table:summary_of_notations_fragment_based}, by replacing all atom types and their covalences in Table \ref{table:summary_of_notations} with fragment types and their number of attachment positions, the whole framework is ready for fragment-based molecular design.

\begin{table}[bp] 
        \caption{Summary of notations used in MIP formulation for fragment-based molecular design. Use the banana odor as an example to provide values for these parameters.}
        \label{table:summary_of_notations_fragment_based}
        \centering
        \vspace{1mm}
        \begin{tabular}{ccc}
            \toprule   
            Symbol & Description & Value \\
            \midrule
            $N$ & number of fragments & $*$ \\
            $F$ & number of features & $14$ \\
            $N^t$ & number of fragment types & $4$ \\
            $N^n$ & number of neighbors & $4$ \\
            $N^h$ & number of hydrogen & $5$ \\
            \midrule
            $I^t$ & indexes for $N^t$ & $\{0,1,2,3\}$ \\
            $I^n$ & indexes for $N^n$ & $\{4,5,6,7\}$ \\
            $I^h$ & indexes for $N^h$ & $\{8,9,10,11,12\}$ \\
            $I^{db}$ & index for double bond & $13$ \\
            \midrule
            $Frag$ & fragment types & fragment set\\ 
            $Cov$ & number of attachment positions & $\{4,2,1,3\}$ \\
            \bottomrule  
        \end{tabular}\\
        fragment set: \{C,O,*C1CCCO1, *C1CCC(*)C(*)C1\}
    \end{table}

This section studies odorant molecules based on the \citet{Sharma2021} dataset.
Within that study, odorant molecules and corresponding olfactory annotations collected from different sources were compiled to a single harmonized resource. The dataset contains $4682$ molecules, spanning over $104$ unique odors, of which two (banana and garlic) were selected for further assessment. Molecules not annotated for a specific odor were assigned the corresponding negative class label. As our formulation focuses on the graph representation without stereo information of molecules, stereo-isomers with contradicting label assignment were removed for the corresponding endpoint. Likewise inorganic compounds or compound mixtures were disregarded in the further analysis. The reasons for testing our approach on both odors are:
\begin{enumerate}
    \item Showing the compatibility of our CAMD formulation: different odors have different groups of fragments and specific chemical interests.
    \item Considering the diversity of the applications: banana is an example of fruity, sweat smell, and garlic represents repelling smells.
\end{enumerate}

The constraints applied so far, i.e., \eqref{C1} -- \eqref{C25}, are mainly of syntactical and very generic nature for theoretic organic molecules. However, there are more specific chemical groups and structures that are unfeasible due to various reasons, including chemical instability, reactivity, synthesizability, or are just unsuited for a desired application. Ideally, the trained GNNs would already capture all those inherent ``rules'' and subtleties. Obviously, this is never the case in a practical setup. In contrast, most interesting applications are located in low data regimes, so the GNNs will often recognize patterns that are important for the applications but might be poor with regard to chemical plausibility. Thus, in a typical de novo design using a model for scoring, a common approach is to add additional constraints based e.g., on SMARTS \citep{SMARTS2007} queries or other simple molecular properties and substructures. This might be done iteratively based on observed structures to additionally add chemical expert knowledge to improve the guidance of de novo design. In the mixed-integer optimization, we can apply additional constraints to narrow down the search space to produce mainly feasible compounds with regard to the desired applications.

\subsection{Case 1: Molecules with banana odor}\label{subsec:banana_odor}
We first fragment all molecules with banana odor in the dataset by cutting all non-aromatic bonds. Since there are only two types of single atoms: C,O, and two types of aromatic rings: *C1CCCO1, *C1CCC(*)C(*)C1, we skip the fragment selection step. The first row of Table \ref{table:banana_features} describes $F=14$ features for each fragment.

    \begin{table}[bp] 
        \caption{Description of fragment features and corresponding coefficients for banana odor. The fragment types are \{C,O,*C1CCCO1, *C1CCC(*)C(*)C1\}. The *=* column denotes that if fragment $v$ is linked with any double bond. Triple bonds are not considered since no molecule with banana odor in the dataset contains triple bond(s).} 
        \label{table:banana_features}
        \centering
        \small\begin{tabular}{c|cccc|cccc|ccccc|c}
            \toprule 
            $X_{v,f}$ & \multicolumn{4}{c|}{fragment type} & \multicolumn{4}{c|}{\# neighbors} & \multicolumn{5}{c|}{\#hydrogen} & *=*\\
            \midrule
            $f$ & 0 & 1 & 2 & 3 & 4 & 5 & 6 & 7 & 8 & 9 & 10 & 11 & 12 & 13\\  
            \midrule
            $h_f$ & $2^{13}$ & $2^{12}$ & $2^{11}$ & $2^{10}$ & $2^{6}$ & $2^{7}$ & $2^{8}$ & $2^{9}$ & $2^{1}$ & $2^{2}$ & $2^{3}$ & $2^{4}$ & $2^{5}$ & $2^{0}$ \\
            \bottomrule  
        \end{tabular}
    \end{table}

To design more reasonable molecules with banana odor, we introduce some chemical requirements to shrink the search space.

\textbf{Requirement 5.1.1} The designed molecule should have at least one of: an aromatic fragment or an oxygen atom linked with double bond. This is based on a visual inspection of molecules with banana odor, where the majority features one of those groups. Mathematically, it requires:
    \begin{align*}
        \exists v\in [N],~s.t.~X_{v,2}+X_{v,3}\ge 1 ~ or ~ X_{v,1}+X_{v,13}\ge 2.
    \end{align*}

To formulate this constraint, we begin by considering the symmetry-breaking constraints \eqref{C26} on feature level:
    \begin{align}\label{B1}\tag{B1}
        \sum\limits_{f\in [F]}h_f\cdot X_{0,f}\le \sum\limits_{f\in [F]} h_f\cdot X_{v,f}+ 2^F\cdot (1-A_{v,v}),~\forall v\in [N]\backslash\{0\},
    \end{align}
where coefficients $h_f,f\in [F]$ are the rearrangement of $2^{F-f-1},f\in [F]$ as shown in the third row of Table \ref{table:banana_features}. The purpose for assigning these coefficients in this specific way is to make sure that if there is either one aromatic fragment or an oxygen atom linked with double bond, its index must be $0$. If fragment $0$ is an oxygen atom, to make sure it is linked with a double bond (which means that it has one neighbor and no hydrogen atom), we can add the following constraint:
    \begin{align}\label{B2}\tag{B2}
        \sum\limits_{f\in [F]}h_f\cdot X_{0,f} \le \underbrace{2^{12}}_{\text{oxygen}}+\underbrace{2^{6}}_{\text{1 neighbor}}+\underbrace{2^{1}}_{\text{0 hydrogen}}+\underbrace{2^{0}}_{\text{double bond}},
    \end{align}
which also includes the cases with at least one aromatic ring.

\subsection{Case 2: Molecules with garlic odor}\label{subsec:garlic_odor}
For molecules with garlic odor, there are four types of atoms: C, N, S, O, and three types of fragments: *C1CCCCC1*, *C1CCC(*)O1, *C1CCSC1. Therefore, there are $F=17$ features for each fragment, i.e., 7 for fragment type, 4 for number of neighbors, 5 for number of hydrogen atoms, and 1 for double bond.

For garlic odor, we propose the following requirements and corresponding constraints.

\textbf{Requirement 5.2.1} Any atom in \{C, N, S, O\} can not be associated with two or more atoms in \{N, S, O\} with single bonds. This requirement is included as the optimization results showed a tendency to include a huge amount of connected heteroatoms in suggested structures. While we might also exclude some chemically feasible molecules with this generic restriction, such structures are often either unstable or reactive. Equivalently, we have:
    \begin{align*}
        \underbrace{\sum\limits_{f=1}^3X_{u,f}}_{u\in \{N,S,O\}}+\underbrace{\sum\limits_{f=0}^3X_{v,f}}_{v\in \{C,N,S,O\}}+\underbrace{\sum\limits_{f=1}^3X_{w,f}}_{w\in\{N,S,O\}}+\underbrace{(A_{u,v}-DB_{u,v})}_{u-v}&+\underbrace{(A_{v,w}-DB_{v,w})}_{v-w}\le 4, \\
        & \forall u\neq v\neq w\in [N].\label{G1}\tag{G1}
    \end{align*}

\textbf{Requirement 5.2.2} Sulfur atom is only linked with carbon atom(s). The reason for introducing this requirement is similar as for Requirement 1, to reduce accumulations of heteroatoms by excluding heteroatom-heteroatom single bonds in the search space. That is, atom S can not be linked with atom(s) in \{N, S, O\}. Similar to \eqref{G2}, we add constraints defined by:
    \begin{align}\label{G2}\tag{G2}
        \underbrace{X_{v,2}}_{v=S}+\underbrace{\sum\limits_{f=1}^3X_{u,f}}_{u\in\{N,S,O\}}+\underbrace{A_{u,v}}_{u-v} \le 2,~\forall u\neq v\in [N].
    \end{align}

\subsection{Extra constraints}
After applying constraints introduced in Section \ref{subsec:banana_odor} and Section \ref{subsec:garlic_odor}, we still find several impractical molecular structures. This section introduces extra constraints for both odors based on different considerations.

\textbf{Requirement 5.3.1} No double bond is linked with an aromatic ring. Since constraints \eqref{C17} already limit the maximal number of double bonds for each fragment, we only need to consider fragments with more than $1$ attachment positions. For banana odor, we exclude cases where fragment *C1CCC(*)C(*)C1 is linked with a double bond, i.e.,
    \begin{align*}\label{B3}\tag{B3}
        X_{v,3}+X_{v,13}\le 1,~\forall v\in [N].
    \end{align*}
For garlic odor, *C1CCCCC1* and *C1CCC(*)O1 should be considered, i.e.,
    \begin{align*}\label{G3}\tag{G3}
        X_{v,4}+X_{v,5}+X_{v,16}\le 1,~\forall v\in [N]. 
    \end{align*}

\textbf{Requirement 5.3.2} Exclude allenes due to their unstability, i.e., a carbon atom associated with two double bonds. This requirement can be implemented by restricting constraints \eqref{C19} into equations:
    \begin{align*}\label{C19'}\tag{C19'}
         X_{v,I^{db}}=\sum\limits_{u\in [N]}DB_{u,v},~\forall v\in [N].
    \end{align*}

\textbf{Requirement 5.3.3} Oxygen atom is not allowed to link with another oxygen atom. We introduce \eqref{B4} to banana odor and \eqref{G4} to garlic odor due to different feature index of oxygen atom, where \eqref{B4} is defined as:
    \begin{align*}\label{B4}\tag{B4}
        \underbrace{X_{v,1}}_{v=O}+\underbrace{X_{u,1}}_{u=O}+\underbrace{A_{u,v}}_{u-v} \le 2,~\forall u\neq v\in [N].
    \end{align*}
and \eqref{G4} is defined as:
    \begin{align*}\label{G4}\tag{G4}
        \underbrace{X_{v,3}}_{v=O}+\underbrace{X_{u,3}}_{u=O}+\underbrace{A_{u,v}}_{u-v} \le 2,~\forall u\neq v\in [N].
    \end{align*}

\textbf{Requirement 5.3.4} We set the maximum number of aromatic rings as $2$ to avoid big molecules, which have low vapor pressure and thus would hardly form gasses which can be smelled. For banana odor, this bound is:
    \begin{align}\label{B5}\tag{B5}
        \sum\limits_{v\in [N]}(X_{v,2}+X_{v,3})\le 2. 
    \end{align}
For garlic odor, this bound is:
    \begin{align}\label{G5}\tag{G5}
        \sum\limits_{v\in [N]}(X_{v,4}+X_{v,5}+X_{v,6})\le 2.
    \end{align}

\section{Numerical results}\label{sec:numerical_results}
GNNs are implemented and trained in PyG \citep{Fey2019}. Mixed-integer formulation of CAMD is implemented based on our extension to OMLT described in Section \ref{sec:GNN_OMLT}. All optimization problems are solved using Gurobi 10.0.1 \citep{Gurobi2023}. All experiments are conducted in the Imperial College London High Performance Computing server. Each optimization problem is solved with a node with AMD EPYC 7742 ($64$ cores and $64$GB memory).

\subsection{Data preparation and model training}
As shown in Section \ref{sec:case_studies}, for each target odor (from banana and garlic) in the dataset from \citet{Sharma2021}, we first fragment all molecules with target odor, then construct features for each fragment and build graph representation on fragment level. After these steps, we have data in the positive class, i.e., with target odor. 

To select data in the negative class, we fragment all molecules without the target odor, and choose molecules consisting of the same set of fragments as the negative class. Since there are many more molecules in the negative class than the positive class, we set the following constraints to filter these molecules:
\begin{itemize}
    \item The upper bound for the number of each type of atom except for C, e.g., N, S, O, is half the number of fragments, i.e., $N/2$.
    \item The upper bound for the number of aromatic rings is $2$.
    \item The upper bound for the number of double bonds is $N/2$.
    \item With banana as the target odor, each molecule should have either one aromatic ring or an oxygen atom linked with double bond.
    \item With garlic as the target odor, each molecule should have either one aromatic ring or a sulfur atom.
\end{itemize}

There are $1278$ molecules consisting of the same group of fragments but without banana odor, $190$ out of which are chosen as the negative class after applying those constraints. Combining with $110$ molecules with banana odor, we have $300$ molecules for the study of banana odor. For garlic odor, $179$ among $1449$ molecules are chosen as the negative class. Together with $83$ molecules in the positive class, we have $262$ molecules for garlic odor.

For each odor, we train a GNN that consists of two SAGEConv layers with $16$ hidden features, a mean pooling, and a dense layer as a final classifier. For statistical consideration, we train $10$ models with different random seeds for each odor. The average training/testing accuracy is $0.905$/$0.703$ for banana odor, and $0.997$/$0.846$ for garlic odor.

\subsection{Optimality: Find optimal molecules}\label{subsec:optimality}
Our optimization goal is to design molecules with the target odor. Denote $y_0$ and $y_1$ as the logits of those trained GNNs. Then the objective function is $y_1-y_0$. Maximizing this objective is equivalent to maximizing the probability of assigning an input molecule to positive class. Together with constraints listed in Section \ref{sec:MIP_CAMD} and Section \ref{sec:case_studies}, our optimization problem is:
    \begin{align*} 
        \max\limits_{(X,A)}~&y_1-y_0\\
        s.t. ~& (y_0,y_1)=GNN(X,A),\\
        ~& \eqref{C1}-\eqref{C25}, \eqref{C19'},\\
        ~& \eqref{C27}, \eqref{B1}-\eqref{B5} \text{ (banana)},\\
        ~& \eqref{C26}, \eqref{C27}, \eqref{G1}-\eqref{G5} \text{ (garlic)}.
    \end{align*}

\begin{figure}[t]
        \begin{subfigure}[b]{0.48\textwidth}
            \centering
            \includegraphics[width=\textwidth]{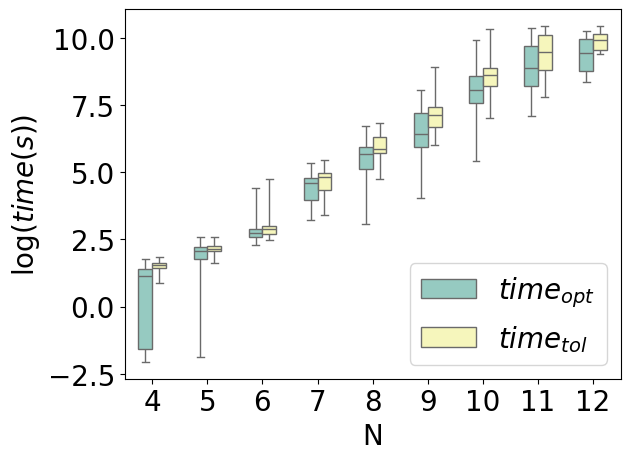}
            \caption{banana odor}
        \end{subfigure}
        \hfill
        \begin{subfigure}[b]{0.48\textwidth}
            \includegraphics[width=\textwidth]{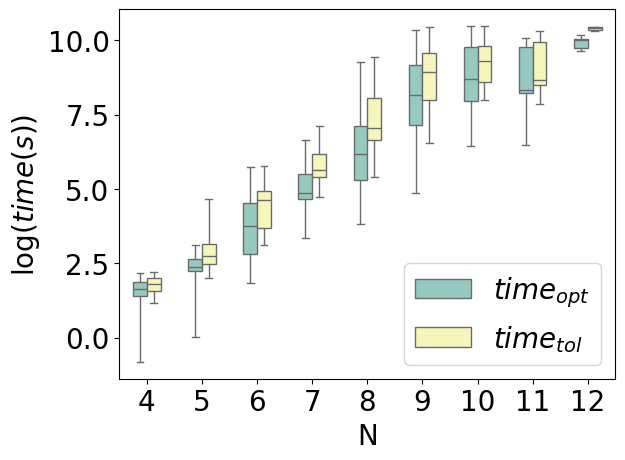}
            \caption{garlic odor}
        \end{subfigure}
        \caption{These graphs report $time_{opt}$ and $time_{tol}$ averaged over the number of successful runs among $50$ runs for both odors. $time_{opt}$ is the first time to find the optimal solution, $time_{tol}$ is the total running time.}
        \label{fig:optimality}
    \end{figure}

\begin{figure}[tp]
        \centering
        \includegraphics[width=.9\textwidth]{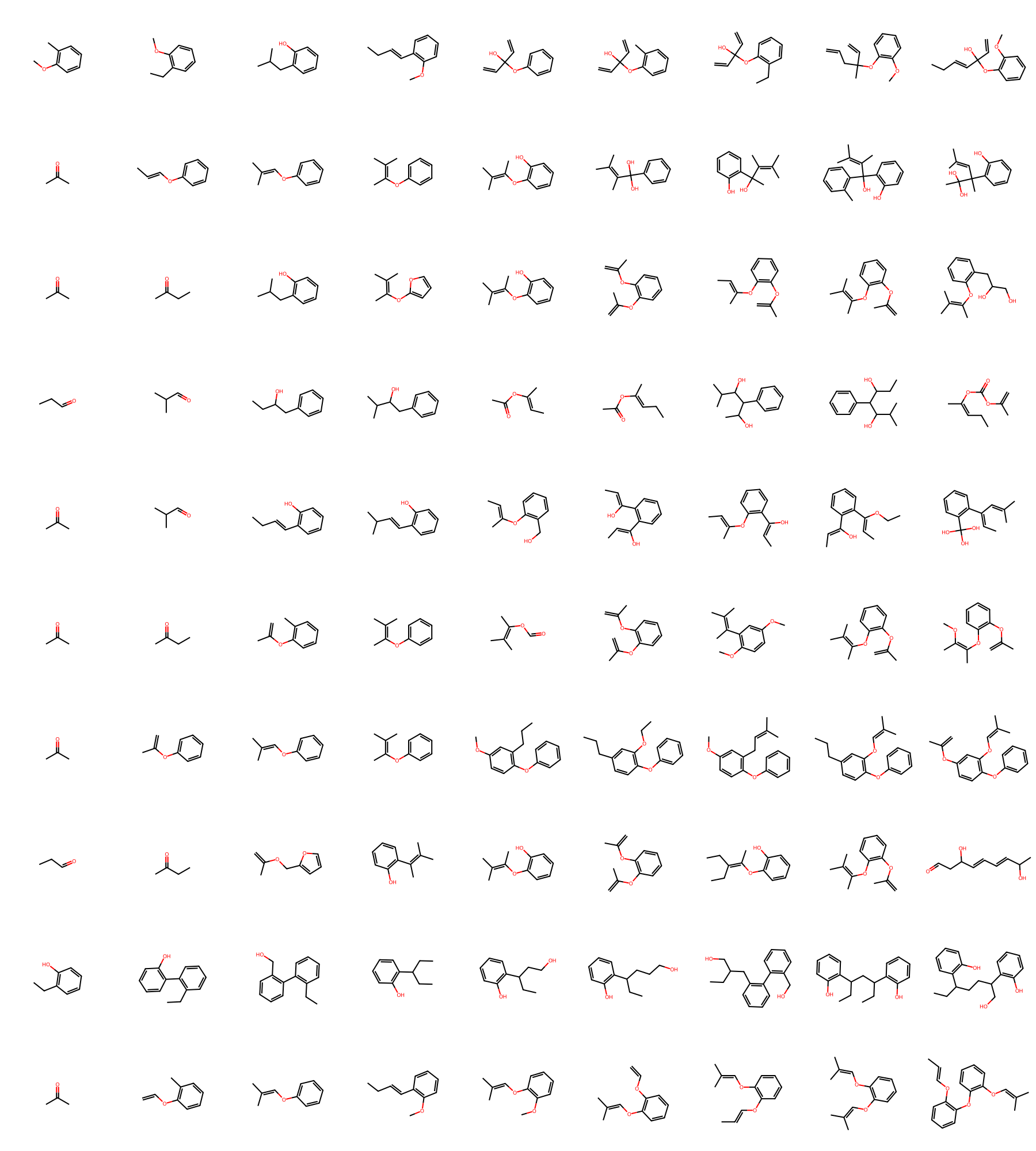}
        \caption{Results for banana odor. Each molecule is the best solution found after $10$ hours for a model with a given $N$. Each row corresponds to the same model, and each column corresponds to the same $N$ ranging from $4$ to $12$.}
        \label{fig:molecules_banana}
    \end{figure}

\begin{figure}
        \centering
        \includegraphics[width=.9\textwidth]{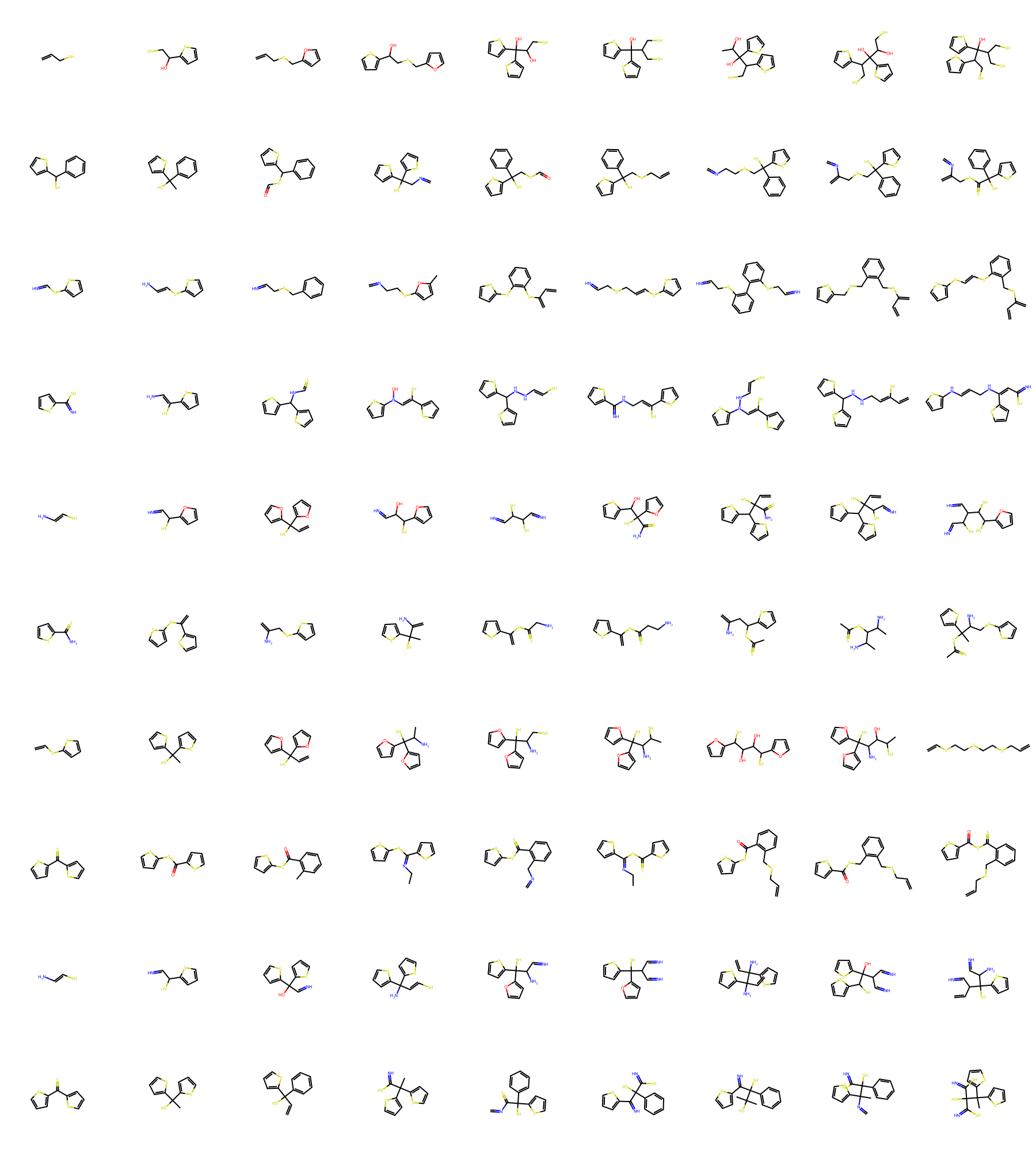}
        \caption{Results for garlic odor. Each molecule is the best solution found after $10$ hours for a model with a given $N$. Each row corresponds to the same model, and each column corresponds to the same $N$ ranging from $4$ to $12$.}
        \label{fig:molecules_garlic}
    \end{figure}

For each model with a given $N$, we solve the corresponding optimization problem $5$ times with different random seeds in Gurobi. Each run uses the default relative MIP optimality gap, i.e., $10^{-4}$, and a $10$ hour time limit. Figure \ref{fig:optimality} shows the time cost for designing molecules with up to $12$ fragments. Figures \ref{fig:molecules_banana} and \ref{fig:molecules_garlic} plot the molecules corresponding to the best solutions found within the time limit. Tables \ref{table:results_banana} and \ref{table:results_garlic} report full experimental details.

    \begin{table}[bp] 
        \setlength\tabcolsep{3pt}
        \caption{Numerical results for banana odor. For each $N$, numbers of variables ($\#var_c$: continuous, $\#var_b$: binary) and constraints ($\#con$) after presolve stage in Gurobi are first reported. Then we count the number of successful runs ($\#run$) over $50$ runs, i.e., achieving relative MIP optimality gap $10^{-4}$. For all successful runs, we provide the $25$th percentile ($Q_1$), mean, and $75$th percentile ($Q_3$) of the running time $time_{tol}$ as well as the first time to find the optimal solution $time_{opt}$. The time limit for each run is $10$ hours. Except for the time  limit and random seed, we use the default setting in Gurobi for other parameters such as the tolerance.}
        \label{table:results_banana}
        \centering
        \vspace{1mm}
        \begin{tabular}{crrrrrrrrrr}
        \toprule
        \multirow{2}{*}{$N$} & \multirow{2}{*}{$\#var_c$} & \multirow{2}{*}{$\#var_b$} & \multirow{2}{*}{$\#con$} & \multirow{2}{*}{$\#run$} & \multicolumn{3}{c}{$time_{opt}$ (s)} & \multicolumn{3}{c}{$time_{tol}$ (s)}\\
        &&&&& mean & $Q1$ & $Q_3$ & mean & $Q1$ & $Q_3$ \\
        \midrule
        $4$ & $478$ & $170$ & $1231$ & $50$ & $3$ & $0$ & $4$ & $5$ & $4$ & $5$ \\
        $5$ & $733$ & $224$ & $1939$ & $50$ & $7$ & $6$ & $9$ & $9$ & $8$ & $10$ \\
        $6$ & $1172$ & $280$ & $2931$ & $50$ & $18$ & $13$ & $18$ & $23$ & $15$ & $20$ \\
        $7$ & $1573$ & $336$ & $3972$ & $50$ & $96$ & $52$ & $121$ & $121$ & $76$ & $146$ \\
        $8$ & $1969$ & $391$ & $5063$ & $50$ & $292$ & $171$ & $374$ & $426$ & $306$ & $546$ \\
        $9$ & $2410$ & $448$ & $6287$ & $50$ & $886$ & $380$ & $1343$ & $1473$ & $792$ & $1679$ \\
        $10$ & $2909$ & $507$ & $7687$ & $50$ & $4204$ & $1990$ & $5291$ & $6237$ & $3736$ & $7104$ \\
        $11$ & $3466$ & $568$ & $9263$ & $41$ & $10686$ & $3684$ & $16495$ & $15954$ & $6815$ & $25038$ \\
        $12$ & $4023$ & $629$ & $10839$ & $9$ & $14869$ & $6461$ & $21331$ & $21415$ & $13923$ & $25480$ \\
        \bottomrule
        \end{tabular}
    \end{table}

    \begin{table}[bp] 
        \setlength\tabcolsep{3pt}
        \caption{Numerical results for garlic odor.}
        \label{table:results_garlic}
        \centering
        \vspace{1mm}
        \begin{tabular}{crrrrrrrrrr}
        \toprule
        \multirow{2}{*}{$N$} & \multirow{2}{*}{$\#var_c$} & \multirow{2}{*}{$\#var_b$} & \multirow{2}{*}{$\#con$} & \multirow{2}{*}{$\#run$} & \multicolumn{3}{c}{$time_{opt}$ (s)} & \multicolumn{3}{c}{$time_{tol}$ (s)}\\
        &&&&& mean & $Q1$ & $Q_3$ & mean & $Q1$ & $Q_3$ \\
        \midrule
        $4$ & $516$ & $187$ & $1363$ & $50$ & $5$ & $4$ & $6$ & $6$ & $5$ & $7$ \\
        $5$ & $811$ & $248$ & $2251$ & $50$ & $12$ & $9$ & $14$ & $21$ & $12$ & $23$ \\
        $6$ & $1315$ & $306$ & $3403$ & $50$ & $65$ & $17$ & $93$ & $108$ & $41$ & $137$ \\
        $7$ & $1759$ & $365$ & $4609$ & $50$ & $214$ & $105$ & $241$ & $383$ & $221$ & $484$ \\
        $8$ & $2204$ & $424$ & $5916$ & $50$ & $1301$ & $198$ & $1247$ & $2397$ & $776$ & $3191$ \\
        $9$ & $2694$ & $484$ & $7345$ & $44$ & $7318$ & $1278$ & $9712$ & $10387$ & $2940$ & $14203$ \\
        $10$ & $3237$ & $546$ & $8962$ & $20$ & $11188$ & $2862$ & $17619$ & $13948$ & $5561$ & $18792$ \\
        $11$ & $3844$ & $609$ & $10790$ & $8$ & $9521$ & $3686$ & $17850$ & $12539$ & $4971$ & $21075$ \\
        $12$ & $4515$ & $674$ & $12833$ & $5$ & $20774$ & $17136$ & $23344$ & $32475$ & $31034$ & $34172$ \\
        \bottomrule
        \end{tabular}
    \end{table}

\subsection{Feasibility: Towards larger design}\label{subsec:feasibility}
As shown in Tables \ref{table:results_banana} and \ref{table:results_garlic}, the increasing dimension and complexity of the optimization problems make solving to optimality intractable with larger $N$. Practically, however, the optimality is not the only priority. A feasible solution with good prediction is also acceptable. In this section, instead of considering optimality, we restrict the time limit to $5$ hours and report the best solutions found within the time limit for larger $N$, i.e., $15\le N\le 50$. By setting \textrm{MIPFocus=1} in Gurobi, we tend to find feasible solutions quickly instead of proving optimality. Also, for each model, we only solve the corresponding optimization problem once due to its large size.

    \begin{figure}[t]
        \begin{subfigure}[b]{0.48\textwidth}
            \centering
            \includegraphics[width=\textwidth]{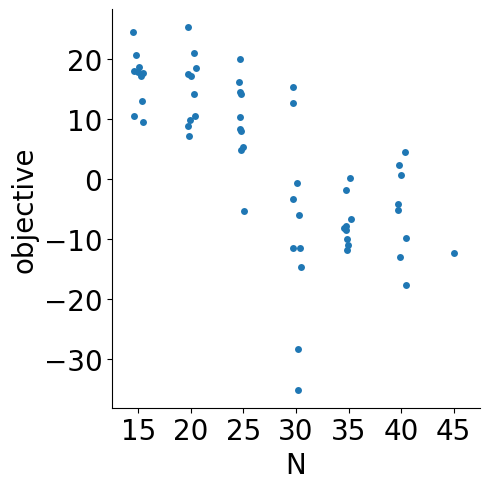}
            \caption{banana odor}
        \end{subfigure}
        \hfill
        \begin{subfigure}[b]{0.48\textwidth}
            \includegraphics[width=\textwidth]{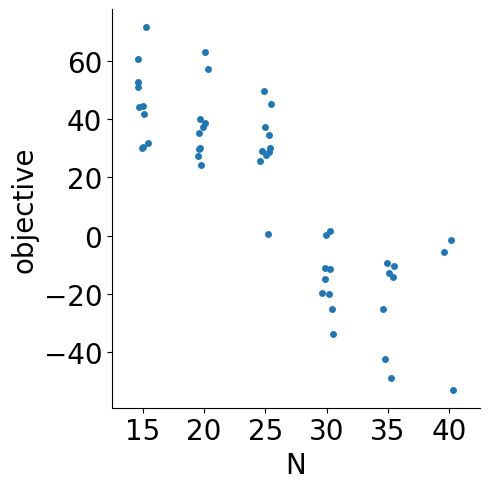}
            \caption{garlic odor}
        \end{subfigure}
        \caption{These graphs report the best objective values found within $5$ hours for $10$ models with $15\le N\le 50$. Each point corresponds to the molecule found from one model.}
        \label{fig:feasibility}
    \end{figure}

As shown in Figure \ref{fig:feasibility}, when $N\le 25$, a positive objective value, i.e., a molecule with target odor, can be found for most runs. When $30\le N\le 40$, however, the solutions are mostly negative, which are unlikely desired odorant molecules. The possible reason is that finding a feasible solution is already hard with such large problem, and there is not enough time for the solver to find better solutions. With even large problems for $N>40$, feasibility becomes impractical: only $1$ run finds a feasible solution among $20$ runs for both odors. For instance, when $N=50$ for garlic odor, there are $9.57 \times 10^4$ variables ($8.82 \times 10^4$ continuous, $7.45\times 10^3$ binary) and $4.67 \times 10^5$ constraints.

\section{Conclusions}\label{sec:conclusion}
This paper considers optimization-based molecular design with GNNs using MIP. To include GNNs into our optimization problems, we propose big-M formulation for GNNs and implement the encodings into open-resource software tool OMLT. To reduce the design space, we introduce symmetry-breaking constraints to remove symmetric solutions caused by graph isomorphism, i.e., one molecule has many different indexing. For case studies, we consider fragment-based design to include aromacity and discover larger molecules. Specific chemical requirements are considered and mathematically represented as linear constraints for both cases. Numerical results show that we can find the optimal molecules with up to $12$ fragments. Without pursuing optimality, our approach can find larger molecules with good predictions.

\section{Acknowledgements}
This work was supported by the Engineering and Physical Sciences Research Council [grant numbers EP/W003317/1], an Imperial College Hans Rausing PhD Scholarship to SZ, and a BASF/RAEng Research Chair in Data-Driven Optimisation to RM.

% %% The Appendices part is started with the command \appendix;
% %% appendix sections are then done as normal sections
% \appendix

% \section{}

 \bibliographystyle{elsarticle-harv} 
 \bibliography{ref}





% \end{thebibliography}
\end{document}